# Coarse Molecular Dynamics of a Peptide Fragment:
# Free Energy, Kinetics, and Long-Time Dynamics Computations

Gerhard Hummer

*Laboratory of Chemical Physics, National Institute of Diabetes and Digestive and Kidney Diseases,*
*National Institutes of Health, Bethesda, Maryland 20892*

Ioannis G. Kevrekidis

*Chemical Engineering, PACM and Mathematics, Princeton University, Princeton, New Jersey 08544*
(Dated: December 12, 2002)

We present a "coarse molecular dynamics" approach and apply it to studying the kinetics and thermodynamics of a peptide fragment dissolved in water. Short bursts of appropriately initialized simulations are used to infer the deterministic and stochastic components of the peptide motion parametrized by an appropriate set of coarse variables. Techniques from traditional numerical analysis (Newton-Raphson, coarse projective integration) are thus enabled; these techniques help analyze important features of the free-energy landscape (coarse transition states, eigenvalues and eigenvectors, transition rates, etc.). Reverse integration of (irreversible) expected coarse variables backward in time can assist escape from free energy minima and trace low-dimensional free energy surfaces. To illustrate the "coarse molecular dynamics" approach, we combine multiple short (0.5-ps) replica simulations to map the free energy surface of the "alanine dipeptide" in water, and to determine the $\sim 1/(1000$ ps) rate of interconversion between the two stable configurational basins corresponding to the $\alpha$-helical and extended minima.

## I. INTRODUCTION

Molecular dynamics (MD) simulations[1] on classical or quantum energy surfaces provide a unique tool for exploring the phase space of (bio)chemical systems at full atomic resolution. In the biological sciences, realistic MD simulations of protein folding, complex formation and aggregation, enzyme kinetics, channel transport, etc., hold the promise to form not only the basis for new understanding of these fundamental processes but also to accelerate the development of new drugs and treatments for diseases. However, these and many other processes in (bio)materials occur on time scales well beyond the reach of current MD simulations, even if carried out with the most powerful computers available on an approximate classical Born-Oppenheimer energy surface. With femtosecond timesteps required to integrate the fastest atomic motions, classical MD of (bio)molecular systems in condensed phase is only starting to push into the microsecond regime,[2,3] and orders of magnitude less on quantum surfaces. Remarkable progress has been made recently in *overcoming* the time scale limitations of MD;[4–9] *circumventing* them is the goal of this work.

The fundamental difficulty in MD arises from the requirement to integrate the motions of "all" nuclear degrees of freedom, not just those of interest. This problem of time-scale separation has been formally addressed in the projection-operator formalism of Zwanzig and Mori[10–12] by constructing a generalized Langevin equation that describes the time evolution in a "slow subspace." The coarse molecular dynamics (CMD) proposed by Kevrekidis and coworkers for equation-free multiscale computations[13–22] (see Ref.[23] for a review) builds on this general framework. The existence of an attracting "slow manifold" for the mesoscopic evolution is assumed. However, explicit constructions of the slow manifold and the corresponding evolution equation, memory kernel, and noise function are avoided by using MD to propagate the microscopic system over intermediate times, as discussed below. Here, we implement the CMD method to study arguably the "simplest biomolecule," alanine dipeptide (i.e., N-acetyl alanine N'-methyl amide), dissolved in water. This system was chosen as a fundamental fragment of protein backbones with torsion degrees of freedom $\varphi$, $\psi$, and $\omega$; and polar groups C=O and N-H that interact strongly with each other and the solvent (see Figure 1). For this reason, the alanine dipeptide has been studied extensively by theory and experiment (see, e.g., Refs. 24–33), and has been used as a model system to analyze the thermodynamics and kinetics of conformational dynamics.[9,34]

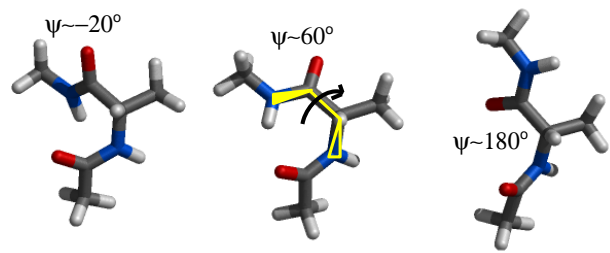

FIG. 1: Structures of the alanine dipeptide in the right-handed $\alpha$-helical minimum (left), in the extended minimum (right), and at the barrier in between (center). The yellow lines and black arrow in the center structure indicate the dihedral angle $\psi$. Carbon, nitrogen, oxygen, and hydrogen atoms are shown in grey, blue, red, and white, respectively.

The two major goals here are (1) to map the essential features of the (coarse) free energy surface of the dipeptide, in particular to find stable minima and connecting saddle points, and (2) to determine the rates of interconversion between the stable minima.

## II. THEORY

### A. Coarse molecular dynamics

In CMD, the existence of an attracting "slow manifold" is assumed. We start with a set of "coarse" variables that *parametrize* this manifold; this means that the slow manifold is the graph of a function over these variables (i.e., it does not "fold" over them). We stress that these variables are "observation variables." They *do not span* the subspace on which the slow dynamics occurs (the "curved" manifold). There exists, however, a one-to-one relation between trajectories on this manifold, and the projections of these trajectories on the hyperplane spanned by the slow variables. The number of such coarse (observation) variables should be at least as large (and preferably, for computational economy purposes, the same) as the dimension of the slow manifold. For macromolecular systems, such "coarse" variables will include descriptors of their internal geometry, such as dihedral angles, radii of gyration, end-to-end distances, number and type of monomer contacts, etc. They may also include solvent coordinates, such as variables describing solvent coordination numbers and structures. In a somewhat different context (e.g., kinetic theory description of fluid flow) the "coarse manifold" will be parametrized by the hydrodynamic variables (density and momentum fields, i.e., low moments of the molecular distribution over velocity space; see for example Ref. 17). If necessary, non-linear couplings between variables can be introduced, for instance, by using the expectations of products of hydrodynamic variables to parametrize the manifold, as in mode-coupling theories. In the CMD approach to atomistic simulations, the assumption is that microscopically evolving distributions in phase space quickly become effectively low-dimensional.[35]

Along fast directions (such as those describing bond vibrations), the distributions rapidly saturate, while they keep drifting and spreading along slow directions (such as those describing large-scale conformational rearrangements). As time progresses, the statistics of motions along the fast degrees of freedom become slaved to the slow degrees of freedom. In a peptide, for instance, the frequency spectrum of fast motions associated with the amide C=O stretch vibration depends on the environment described by the slow variables quantifying, e.g., the type and amount of secondary structure and solvent present. Geometrically, the ensemble-averaged trajectories relax onto a low-dimensional, attracting, forward invariant manifold in coarse phase space. This picture invokes an analogy with the so-called "inertial manifolds" for dissipative partial differential equations.[36–38]

The (expected) dynamics of the coarse variables are explored through short multiple replica simulations as follows (see Figure 2): A coarse initial condition is prescribed, and "lifted" to many microscopic copies consistent with (conditioned on) the coarse variables. This "lifting" step is not unique, since many distributions can be constructed that have the same coarse variables. Lifting can be achieved, for in-stance, by performing a short MD run with an added potential biasing the coarse variables toward their new target values, as in umbrella sampling.[39] This approach works also for coarse variables consisting of nonlinear combinations of the atomic coordinates, as in the product variables of mode-coupling theories. The initialization will necessarily be at least slightly "off manifold;" that is, the fast degrees will not be initialized to be the "correct" functionals of the coarse variables. This discrepancy, however, will be quickly "healed," as a relatively short detailed simulation will bring the trajectory down on the manifold (i.e., slave the statistics of the fast degrees of freedom to the coarse variables). After this short healing or "maturing" period we monitor the detailed simulations over short times, and estimate the time derivative of the slow dynamics in the coarse variables. This time-derivative can be used to extrapolate coarse dynamics over relatively long time intervals; repeating the procedure (lift, short run, restrict to coarse variables, estimate coarse variable time derivative, project coarse variables into the future, lift again . . . ) constitutes the coarse projective integration schemes of Gear and Kevrekidis,[15,19,40] and allows us to extrapolate the long-time coarse behavior from short, repeated bursts of appropriately initialized MD simulation. It is important to stress the relation between these methods and the optimal predictors of Chorin and coworkers.[41,42] Beyond these "accelerated" dynamics, we will see that this approach permits the rapid search of dynamically-relevant features of the free energy surface and the calculation of transition rates, as discussed below. Here, we use the dihedral angle $\psi$ (N-C$_\alpha$-C-N) of the Ala dipeptide as a first approximation for the coarse coordinates (i.e., the coordinates used to observe the slow dynamics). In the following, we will explain the method in terms of this coordinate. Generalizations to multiple dimensions are straightforward, have been discussed in the literature,[18,22,23] and will be illustrated here by occasionally using the dihedral angle $\varphi$ (C-N-C$_\alpha$-C) as a second coarse coordinate.

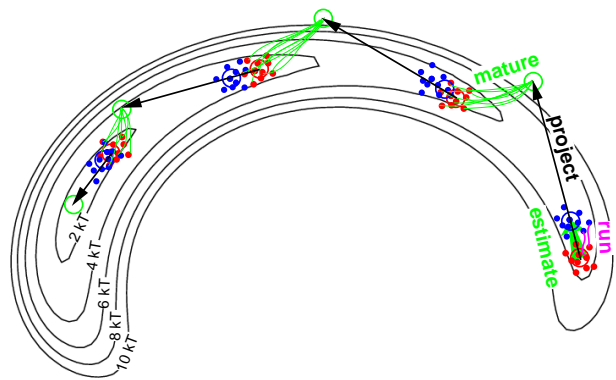

FIG. 2: Schematic representation of projective forward integration in CMD for a two-dimensional coarse free energy surface. Short MD runs (magenta lines) connect multiple nearby initial (red dots) and final configurations (blue dots). From the drift in the average coarse variables (red and blue circles), a forward time derivative is estimated which is then used in a projective forward step (black arrow). *Lifting* (e.g., by using umbrella sampling) produces one or several configurations at the projected position, from which multiple replica runs are initialized. After maturing (thin green lines) by relaxing on the free energy surface, the estimated drift in the average coarse variables is used for another forward projection. In this schematic plot, "convergence" to the free energy minimum is practically achieved in four steps.

### B. From drift in the coarse variables to free energies

Assume that we have created an ensemble of configurations $i$ with identical values of $\psi_i(t = 0; \psi_0) = \psi_0$, for instance by randomly assigning Maxwell-Boltzmann velocities to all solute and solvent atoms in a given configuration. As these configurations evolve in time, the dynamics of $\psi_i(t; \psi_0)$ will initially be governed by the coupling to fast motions such as bond vibrations and molecular collisions. As a consequence, the $\psi_i(t; \psi_0)$ will "spread" rapidly over a small range. After a sufficiently long time $t > \tau_{\text{mol}}$, however, the average $\overline{\psi(t; \psi_0)}$ curve is expected to drift down on the free energy surface (or potential of mean force) $G(\psi)$ towards a stable minimum. "Timestepping" for $\overline{\psi(t; \psi_0)}$ involves lifting, evolving for a time $t > \tau_{\text{mol}}$, and restricting back to coarse variables (i.e., collapsing the distribution of the endpoints of the several trajectories to their average). This "coarse timestepper" (the map resulting from the lift-run-restrict procedure), lies at the heart of the coarse numerical processes (coarse integration, coarse Newton-Raphson, coarse optimization) we perform. Initializing at nearby values of $\psi_0$ can be used to estimate the partial derivatives of the coarse timestepper with respect to its (coarse) initial conditions, and from that we can obtain estimates of the local curvature of the free energy surface. From a Langevin approximation to the dynamics of slow variables,[12] we expect that the drift velocity is proportional to the slope of the free-energy surface:

$$\frac{\partial \overline{\psi(t; \psi_0)}}{\partial t} \approx -D\beta \frac{\partial G(\psi)}{\partial \psi} \quad \text{for } t > \tau_{\text{mol}} \qquad (1)$$

where $\beta^{-1} = k_B T$ (with $k_B$ Boltzmann's constant, $T$ the temperature) and $\psi = t^{-1} \int_{t_0}^{t} \overline{\psi(t'; \psi_0)} dt'$. Normally, one would begin averaging after an initial maturation time ($t_0 > \tau_{\text{mol}}$); here, we averaged over the whole interval ($t_0 = 0$). The proportionality constant $\beta D$ is a mobility, with $D$ a diffusion coefficient that may depend on $\psi$. We may also expect that the variance in $\psi_i(t; \psi_0)$ of different runs $i$ initialized with the same $\psi_i(0; \psi_0) = \psi_0$ will exhibit diffusive behavior,

$$\frac{\partial}{\partial t} \text{var}[\psi(t; \psi_0)] \approx 2D(\psi) \qquad (2)$$

where the diffusion coefficient $D$ is related to the mobility in Eq. (1) through the Einstein relation. We expect Eq. (2) to hold for times $\tau_{\text{mol}} < t \ll (D|\partial^2 G/\partial \psi^2|)^{-1}$ long compared to the initial molecular relaxation time $\tau_{\text{mol}}$ but short with respect to curvature effects of the free energy surface. One can thus use the time-evolution observed through the coarse variables to estimate the deterministic and stochastic components of the motion, and from those estimate the underlying free energy surface, $G(\psi)$, and dynamic properties, such as effective diffusion coefficients $D(\psi)$.

### C. Finding stationary points on the coarse free energy surface

In many practical applications, one is interested in finding the stable minima and the (unstable) saddle points on the coarse free energy surface, rather than mapping the complete surface. Protein folding simulations constitute one such example with goals of finding the folded structure and the structures at the "barrier" to folding. From Eq. (1), we expect that the minima and maxima are stable and unstable stationary points for the drift component of the dynamics in the coarse variables. We can thus use the dynamic information to search for these stationary points on the free energy surface, i.e., minima and saddles satisfying

$$\frac{\partial \overline{\psi(t; \psi_0)}}{\partial t} = 0 \quad \text{or } \psi_0 - \overline{\psi(t; \psi_0)} = 0. \qquad (3)$$

By estimating $\overline{\dot{\psi}(t; \psi_0)} = \partial \overline{\psi(t; \psi_0)}/\partial t$ or $\psi_0 - \overline{\psi(t; \psi_0)}$ for nearby initial points $\psi_0$, we can estimate $\partial \overline{\dot{\psi}}/\partial \psi$ and perform a Newton-Raphson type step towards a stationary point:

$$\psi_1 = \psi_0 - \frac{\overline{\dot{\psi}}}{\partial \overline{\dot{\psi}}/\partial \psi_0} \quad \text{or } \psi_1 = -a/b \qquad (4)$$

where $a$ and $b$ are the intercept and slope, respectively, in a straight-line fit of $\psi_0 - \overline{\psi(t; \psi_0)}$ versus $[\psi_0 + \overline{\psi(t; \psi_0)}]/2$ for nearby starting values of $\psi_0$. Recursive application of Eq. (4), can be used to converge to a stationary point. This timestepper approach to coarse steady state computations was introduced for spatially distributed systems,[13] and was illustrated for stochastic systems[18,20,22] (stable and saddle stationary states of kinetic Monte Carlo and Brownian Dynamics simulations).

Unlike the thermodynamically determined free energy profiles (e.g., from umbrella sampling[39] or constrained dynamics[43]) the dynamic sampling also gives immediate indication about the kinetic relevance of the chosen coarse coordinate. Consider the classic example (see, e.g., Figure 1b in Ref. 44) of a rotated double well system in two dimensions. Projected onto the $x$ axis, the two wells overlap in the barrier region. If the dynamics is monitored only along the $x$ axis but not the "solvent" coordinate $y$, most of the structures at the barrier in $x$ appear already "committed" to one or the other well. This has important implications. In free energy calculations using umbrella sampling or thermodynamic integration, the system will not be pulled easily over the barrier by a potential acting on $x$ alone; in kinetics calculations, the dynamics along $x$ alone is non-Markovian, making rate calculations difficult. In our dynamic analysis, this situation becomes manifest in a large spread of the estimated drift velocity and direction, $\overline{\psi(t; \psi_0)}$, for nearby initial values of $\psi_0$.

A detailed discussion of practical approaches to determining the need for additional coarse variables is contained in Refs. 18 and 22. The idea is to quantify this need "on-the-fly" either indirectly (through the concomitant gradual loss of precision in estimating coarse derivatives) or directly (lifting with more coarse variables, determining the eigenvalues of the local linearization, and testing whether a gap remains, or whether an additional eigenmode is becoming slow). The eigenvector of this local eigenmode will then suggest a good additional coarse variable or combination of variables for the augmentation. These approaches very much resemble, in spirit, tests for the adaptive mesh refinement or adaptive step size selection in partial-differential-equation simulations: there we compare the results using smaller (or larger) timesteps and finer (or coarser) meshes and decide whether to refine (or coarsen) the mesh. Here, in a completely analogous way, we are pursuing the "adaptive coarse variable augmentation" (and hopefully, in some cases, reduction!) for CMD (and other multiscale) codes.

From the CMD analysis, we can also extract immediately properties of the dynamics in the free energy wells. In particular, we can estimate the correlation time $\tau_c$ for slow (diffusive) motion at the bottom of a well from the slope of the drift velocity with respect to the coarse variable:

$$\tau_c^{-1} = -\left.\frac{\partial \overline{\dot{\psi}}}{\partial \psi}\right|_{\overline{\dot{\psi}}=0} \qquad (5)$$

The local correlation time $\tau_c$ in a particular steady state can be determined directly from the Newton-Raphson search upon convergence. The slope of $\overline{\dot{\psi}}$ with respect to $\psi$ provides the "restoring force" for the coarse variable. If $\overline{\dot{\psi}}$ remained linear in $\psi$, then the motion would correspond to a harmonic oscillator in the high-friction (overdamped) limit. In many dimensions, the characteristic frequencies are given by the eigenvalues of the Jacobian analogous to the derivative in Eq. (5). The corresponding eigenvectors give the "normal coordinates" for the overdamped motions around a stationary point.[45]

### D. Coarse projective integration

The dynamic information in the replica runs $\psi_i(t; \psi_0)$ can also be used to extrapolate toward longer times. Instead of propagating each of the replicas, we extrapolate the *average* position of the slow variable, for instance linearly (exploiting regularity of the expected coarse dynamics with time):

$$\overline{\psi(t'; \psi_0)} \approx \overline{\psi(t; \psi_0)} + \frac{t'-t}{t}\left[\overline{\psi(t; \psi_0)} - \psi_0\right] \qquad (6)$$

A long "projective" step $t'-t$ is then effected by re-initializing an ensemble at the extrapolated value; this is the simplest "Projective Forward Euler Method."[15] Clearly, instead of simply taking a linear interpolation between the first and last coarse values of a short run, we can record a short "tail" of the coarse evolution after the quick maturing period and use that to construct (linear or higher order) predictors.[15,19,40]

In the Ala dipeptide, we use a harmonic constraint on the torsion potential to initialize $\psi$ at its new target value. With new initial velocities assigned from a Maxwell-Boltzmann distribution, the fast degrees of freedom, such as bonds or solvent positions, re-equilibrate rapidly and the newly initialized state starts drifting again on the free-energy surface. It is interesting that in addition to coarse projective integration and to coarse Newton-Raphson, additional algorithms, like time-stepper based coarse control[21] and coarse optimization[46], become enabled, and they may be useful in analyzing these surfaces without ever explicitly closing Langevin equations on them.

### E. Reverse coarse integration and escaping free energy minima

It is well known that whether one uses MD forward or backward in time (i.e., whether one flips the velocities of the initial configuration or not) one obtains *forward* evolution of the coarse variables. As discussed in Ref. 47, however, it is possible to integrate *the coarse variables* backward in time on the slow manifold (i.e., on the coarse free energy surface) exploiting projective integration as follows. After initializing at the molecular level, and running (whether with flipped or unflipped velocities) long enough for the lifting errors to heal, we use the estimated coarse *forward* time derivative $\overline{\dot{\psi}}$ or $\psi_0 - \overline{\psi(t; \psi_0)}$ to perform a relatively large *reverse* projection in time. Under appropriate conditions this is a stable algorithm.[48]

For reverse integration from $\psi_0$ equal to the length of the forward healing/estimation step, we can use a simple Euler integrator:

$$\overline{\psi(-\Delta t; \psi_0)} \approx 2\overline{\psi(0; \psi_0)} - \overline{\psi(\Delta t; \psi_0)} \qquad (7)$$

We then (1) lift again from the projected $\psi(-\Delta t; \psi_0)$, (2) run again MD (whether forward or backward in time) to obtain a new estimate of the local, coarse forward-in-time $\overline{\dot{\psi}}$ or $\psi_0 - \overline{\psi(t; \psi_0)}$; and (3) reverse project again the coarse variable backward in time.

What was discussed here is the simplest "projective Forward Euler" method used backward in time; it is clear that more elaborate multistep coarse projective integration methods can be used to accelerate both the forward and reverse coarse integration in our context. Notice that this "reverse integration" is useful not only in a microscopic/stochastic context, but also in the case of stiff deterministic problems in general, and even discretizations of dissipative partial differential equations[48]. Such generalizations of coarse projective integrations (including coarse projective Runge-Kutta, Adams, and even implicit implementations of such algorithms in the "coarse" case) are outlined in Refs. 15 and 40 and are the subject of ongoing research in collaboration with C. W. Gear. This "short step forward in the full space, large step backward approximately on the manifold" can be used to systematically integrate the unavailable equations backward in time on the free energy surface. For this one-dimensional problem, the "saddle" is actually a "source," i.e., an attractor *backward* in time. Suitably initialized, reverse integration of $\overline{\psi(t; \psi_0)}$ will then converge to the saddle where $\overline{\dot{\psi}(t; \psi_0)} = 0$ and $\partial \overline{\dot{\psi}(t; \psi_0)}/\partial \psi > 0$. Reverse integration can become an efficient way of both exploring the surface and escaping free energy minima. It appears that, if all but one or two of the coarse backward directions are very stiff (i.e., a separation of time scales prevails for the coarse variable drift), their effect is quickly damped by the short forward integrations. With the appropriate step choices, the reverse integration will probably only "see" the slow backward directions, and use them to "climb back up" the slow backward path(s). Methods for the construction of stable manifolds for low-dimensional dynamical systems (see, e.g., Refs. 49,50) can, under favorable time scale separation conditions, be brought to bear on the computational endeavor of exploring the free energy surface through reverse coarse integration.

## F. Calculation of rates

We can also use CMD to estimate rates of interconversion between states. Instead of performing one long run, we determine the short-time dynamics in the space spanned by the coarse variables from many short, appropriately initialized replica runs. The replica runs are used to construct propagators. Assuming some degree of regularity (smoothness) these propagators are then applied recursively to infer the long-time dynamics. From $N$ replica simulations of length $t$, we can estimate a propagator as a sum of $\delta$ functions at the end points of each trajectory:

$$p(\psi, t|\psi_0, 0) \approx N^{-1} \sum_{i=1}^{N} \delta[\psi - \psi_i(t; \psi_0)] \qquad (8)$$

Alternatively, propagators could be built by filtering the data or by using cumulants of the distribution at time $t$. Under the assumption of Markovian dynamics, the distribution at times of $2t$, $3t$, etc., can be determined by recursive application of the Chapman-Kolmogorov identity,

$$p(\psi, 2t|\psi_0, 0) = h(\psi) \int p(\psi, t|\psi', 0)p(\psi', t|\psi_0, 0)d\psi' \quad (9)$$

Without absorbing points [$h(\psi) = 1$ for all $\psi$], we expect to recover the equilibrium distribution by iterating to infinite time, $\beta G(\psi) = -\ln p(\psi, t \to \infty|\psi_0, 0)$. Equation (9) provides an alternative route to the free energy surface, in addition to Eq. (1). We note that under the Markovian assumption, the action functional describing the relative probabilities of a dynamic path (here: $\psi_0$, $\psi_1$, ..., $\psi_N$) in the space of the coarse variables is given by the product of the corresponding propagators:

$$e^{-s(\psi_0, \psi_1, \ldots, \psi_N)} \equiv e^{-\beta G(\psi_0)} \prod_{i=0}^{N-1} p(\psi_{i+1}, t|\psi_i, 0) \quad (10)$$

We can use Eq. (9) to estimate "reaction rates." To find first-passage time distributions from a "reactant" state to a "product" state, absorbing points can be inserted by multiplying the integral in the Chapman-Kolmogorov relation, Eq. (9), with $h(\psi) = 0$ inside the "product" region and $h(\psi) = 1$ outside. This assumes that the time $t$ is short relative to the average time for escape from the reactant and product well. Integration of $p(\psi, nt|\psi_0, 0)$ over $\psi$ then gives a survival time distribution at time $nt$ starting from $\psi = \psi_0$:

$$S(nt) = \int p(\psi, nt|\psi_0, 0)d\psi \qquad (11)$$

Application of Eq. (9) requires propagators at intermediate $\psi$ values. This is where regularity in coarse phase space is assumed: propagators at intermediate values can be estimated by interpolation from replica runs initialized with different values of $\psi$. For an intermediate value $\psi_2$, we can use a simple linear interpolation:

$$p(\psi, t|\psi_2, 0) \approx (1 - \alpha)p(\psi, t|\psi_0, 0) + \alpha p(\psi, t|\psi_1, 0) \quad (12)$$

or

$$\begin{aligned} p(\psi, t|\psi_2, 0) \approx\ & (1 - \alpha)p(\psi + \psi_2 - \psi_0, t|\psi_0, 0) \\ & + \alpha p(\psi + \psi_2 - \psi_1, t|\psi_1, 0) \end{aligned} \quad (13)$$

where $\alpha = (\psi_2 - \psi_0)/(\psi_1 - \psi_0)$ with propagators for $\psi_0$ and $\psi_1$ given by Eq. (8). For diffusion on a linear potential with $D$ constant, the linear interpolation Eq. (13) is exact. If the diffusion coefficient is position dependent and possibly anisotropic, appropriate generalizations can be constructed. (Notice that our CMD approach *does* provide estimates of the position dependence and the anisotropy of the diffusion coefficient.)

In a certain limit, the concept of an optimal path can be helpful in rate computations. Optimization *over coarse variable paths* can be performed using CMD timesteppers. We can locally approximate both the deterministic *and* the "diffusive" part of the evolution through the propagators estimated from replica runs. If we guess a coarse transition path between two coarse minima we can write down the coarse action functional for this path. The coarse action functional can thus be deterministically minimized, using, for example, dynamic programming methods. One can estimate the sensitivity of the functional to (discretized) path variations through nearby short integrations; or, alternatively, "derivative free" optimization methods can be used for this goal, and the conversion rate computed (approximated) upon convergence of the optimization.[51] We have, in the past, solved discretized coarse optimization problems for optimal parameter variation policies;[52] we are currently pursuing the application of the same optimization methods for the computation of rates in problems with coarse dimension higher than one.

In many cases, the concept of the optimal path loses its significance in the computation of rates. Indeed, one now evolves a state density over the coarse free energy surface, solving, in effect, a Fokker-Planck equation for the evolution of this density and its stationary state. For a relatively low coarse-dimensional problem (i.e., if the coarse free energy surface is two- or three-dimensional) it should be possible to solve this Fokker-Planck by convolving the propagators above.

In summary, we force the system to sample the dynamics in weakly populated regions of the "coarse variable space" through multiple initializations of the coarse variables. This allows us to circumvent the problem that the short-time dynamics "fully" samples only the fast degrees of motion (as conditioned on the slow "coarse variables"), but covers only a small range in the slow variables. We are therefore trading multiple initializations for long time dynamics. This allows us to sample even the slow dynamics in the coarse variables. This can be a computationally efficient approach, as can be understood for 1D diffusive barrier crossing. By constructing and interpolating the propagators along the coordinate leading across the barrier, the rate of crossing can be estimated for arbitrarily high barriers. It is important to state, however, that this procedure assumes the dynamics to be "smooth," such that propagators can be interpolated, and that the dynamics in the coarse variables is Markovian at the time scale of the short replica simulations.

Interpolating the propagators to solve a Fokker-Planck equation brings up an analogy with the so-called "gap-tooth" methods in Refs. 14,23. In this approach, regularity of the solution of a Fokker-Planck equation in space and time can, in principle, be used to accelerate a particle-based solver by evolving particles in "patches" of space-time, separated by empty gaps. Communication across the "gaps" of the "teeth" in which the particle density evolves is, of course, the key to the approach. The construction of successful boundary conditions capable of effecting this may significantly enhance the performance of Brownian Dynamics type solvers[23,53,54].

## III. MOLECULAR DYNAMICS SIMULATIONS

MD simulations of the hydrated Ala dipeptide are performed with the sander module in the AMBER 6.0 simulation package (University of California at San Francisco) and the parm94 force field.[55] The dipeptide is simulated in a periodically replicated box with 607 TIP3P water molecules.[56] Particle-mesh Ewald summation is used for the long-range electrostatic interactions.[57] The system is simulated at constant volume corresponding to $\sim 1$ bar pressure, as determined during an initial equilibration run. The temperature is maintained at 300 K by weak coupling (10 ps time constant) to a Berendsen thermostat.[58] Bond lengths involving hydrogen atoms are constrained using the SHAKE algorithm.[59] After 500 ps of equilibration, we collect data for 7 ns, with configurations saved every 0.5 ps for analysis. All simulations use a time step of 0.001 ps.

Replica runs are initialized by drawing particle velocities from a Maxwell-Boltzmann distribution. For rigid TIP3P water, rigid body translational and rotational velocities are generated in the principal-axes system. Initial velocities along (possibly linked) constrained bonds of the dipeptide are removed recursively.

"Lifting" to a new target value of the coarse coordinate $\psi$ in CMD is accomplished by a short run with a tight harmonic potential, $k(\psi - \psi_0)^2/2$, acting on the dihedral angle $\psi$, with $k = 100$ kcal mol$^{-1}$ rad$^{-1}$. With "lifting" here occurring not instantaneously but over a finite time interval, the fast variables are given time to adapt to the changes in the coarse variable. This effectively shortens the time needed to mature the system at the new state, as prescribed by the updated coarse variable.

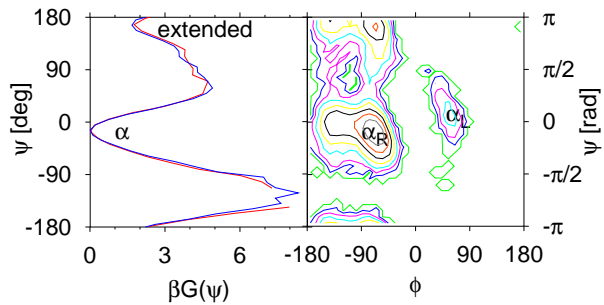

FIG. 3: Free energy surfaces of the alanine dipeptide. (Right) Free energy surface in the $\varphi$-$\psi$ plane from the 24-ns run with 265 water molecules ($1 k_B T$ contour lines). (Left) Free energy in units of $k_B T$ (horizontal axis) as a function of the $\psi$ dihedral angle (vertical axis). The red and blue lines are the results from the 7 and 24-ns runs with 607 and 265 water molecules, respectively.

At every iteration of the Newton-Raphson search for the location of stationary points, ten initial configurations are created as structures along a 5-ps run with a harmonic constraint holding $\psi$ near the target value. For each of the ten configurations, ten sets of random initial velocities are drawn from a Maxwell-Boltzmann distribution, followed by 0.5 ps of MD.

## IV. RESULTS AND DISCUSSION

### A. Equilibrium molecular dynamics

From 7 and 24-ns equilibrium runs with 607 and 265 water molecules, respectively, we estimate the free energy surface $\beta G(\psi) = -\ln p(\psi)$ shown in Figure 3, where $p(\psi)$ is given by the histogram of the dihedral angle $\psi$. We find two minima, one corresponding to a right-handed $\alpha$ helix [with $G(\psi \approx -0.3$ rad$) \approx 0$] and the other to an extended structure [with $G(\psi = \pi) \approx 1$ kcal mol$^{-1}$]. In the two-dimensional $\varphi$-$\psi$ Ramachandran plot of Figure 3, we observe a small population in the left-handed $\alpha$-helical minimum during the 24-ns run, but not the 7-ns run. In the following, we will focus on the equilibrium between the extended and right-handed $\alpha$ helical structures. The lower of the two barriers separating the two minima, with a height of about 3 kcal mol$^{-1}$ near $\psi = 1$ rad, shows a considerable amount of structure, with a small dip (7-ns run) or shoulder (24-ns run) near the polyproline $P_{II}$ minimum ($\psi \approx 2\pi/3$).

For the second barrier near $\psi = -2\pi/3$ rad, we estimate a height of about 5 kcal mol$^{-1}$ from umbrella sampling in the barrier region, in agreement with the value obtained by Bolhuis et al.[34] Overall, our equilibrium runs give a very similar $G(\psi)$ to the one obtained by Bolhuis et al.[34] with umbrella sampling.

From the variance var$_\alpha(\psi)$ of $\psi$ in the $\alpha$-helical minimum, and the decay time $\tau_c$ of the corresponding autocorrelation function, we estimate a diffusion coefficient of $D \approx$

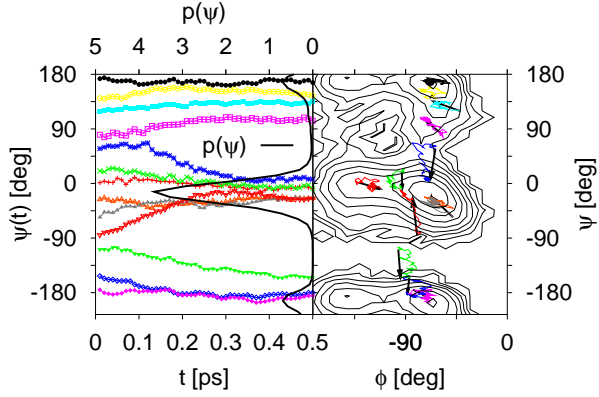

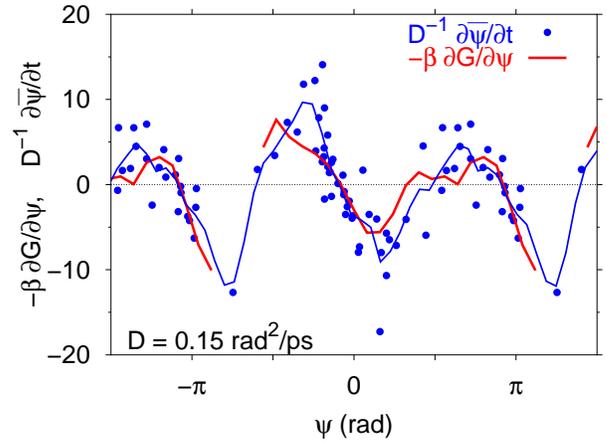

FIG. 4: Drift in the $\psi$ dihedral angle. The panel on the left shows $\overline{\psi(t;\psi_0)}$ as a function of time averaged over 50 runs with different initial velocities. The thick black line shows the equilibrium probability (horizontal axis; arbitrary units) of $\psi$ (vertical axis) to demonstrate that $\overline{\psi(t;\psi_0)}$ converges toward the most populated regions of $\psi$. The panel on the right shows the "drift" of $\overline{\varphi(t;\varphi_0)}$ and $\overline{\psi(t;\psi_0)}$ in a Ramachandran plot (colored lines). Initial configurations of the replica runs were chosen from structures along the 7-ns equilibrium run. Thick arrows point from the initial values of $\varphi$ and $\psi$ to the final values. Thin black lines are free energy contours separated by 1 $k_B T$.

$$\text{var}_\alpha(\psi)/\tau_c \approx 0.15\ \text{rad}^2\ \text{ps}^{-1}.$$

### B. Free energy surface

To map the free energy surfaces with CMD, we assume that the distributions of fast variables (bond lengths, etc.) are quickly slaved to the slow variables, and that the averaged dynamics of coarse variables is governed by the underlying free energy surface. We thus expect that the averaged projections onto the coarse variables drift over a short (here: subps) timescale towards free energy minima. Figure 4 shows $\overline{\psi(t;\psi_0)}$ as a function of time $t$ for runs starting from 13 different initial configurations.

Also shown is the probability distribution $p(\psi)$ of $\psi$ from the 24-ns equilibrium run (with $\psi$ on the vertical axis). We find that $\overline{\psi(t;\psi_0)}$ indeed drifts toward the maxima of $p(\psi)$ and away from minima. Similar behavior is found for the corresponding dynamics projected onto the $\varphi$-$\psi$ Ramachandran plane. To get a more quantitative estimate of the free energy surface, we estimate $\overline{\dot\psi}$ throughout the interval $-\pi < \psi < \pi$. From 67 different initial configurations, we run 50 short ($\tau = 0.5$ ps) replica simulations. The $\psi_i(t;\psi_0)$ curves of the replicas are averaged and fitted to straight lines. Figure 5 shows the corresponding slope $\overline{\dot\psi}$, scaled by $D = 0.15\ \text{rad}^2\ \text{ps}^{-1}$, as a function of the average angle $\langle\overline\psi\rangle = \tau^{-1}\int_0^\tau \overline{\psi(t;\psi_0)}dt$. As mentioned before, we average here over the whole interval, including the initial maturation time. We also calculated the local diffusion coefficient $D(\psi)$ from Eq. (2). Near the minima of $G(\psi)$, we recover the value for the equilibrium run. Near barriers, however, curvature effects are relevant even at the 0.5 ps timescale because of the relatively rapid relaxation of $\psi$ compared to $\tau_{\text{mol}}$. For simplicity, we will in the following use a constant $D = 0.15\ \text{rad}^2\ \text{ps}^{-1}$ instead of correcting $D(\psi)$ for the curvature effects by using the $\psi_0$-dependence of the $\overline{\psi(t;\psi_0)}$ data.

FIG. 5: Drift velocity $\overline{\dot\psi}$ as a function of the $\psi$ dihedral angle. Shown are $D^{-1}\overline{\dot\psi}$ (blue symbols) and the mean force $-\beta\partial G(\psi)/\partial\psi$ (red line) for $D = 0.15\ \text{rad}^2\ \text{ps}^{-1}$. The smooth blue line is a spline approximation to the $D^{-1}\overline{\dot\psi}$ data.

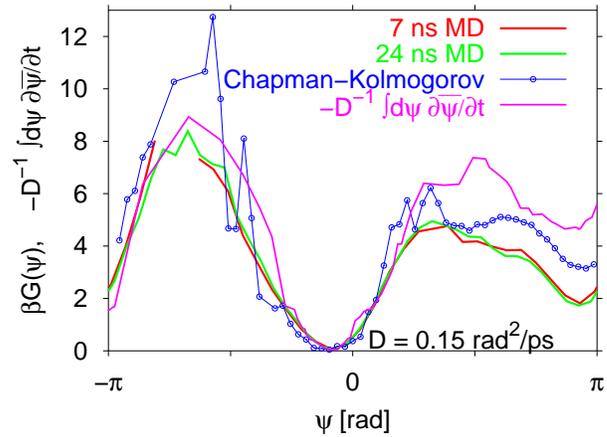

FIG. 6: Free energy surface as a function of $\psi$. Results are shown for the two equilibrium runs with 265 (green line) and 607 water molecules (red line), and for integration of $D^{-1}\overline{\dot\psi}$ (magenta line). Also shown is the free energy corresponding to the limiting distribution of Chapman-Kolmogorov iterations (blue line with symbols), evaluated at 56 discrete values of $\psi$.

As expected from Eq. (1), the $D^{-1}\overline{\dot\psi}$ data scatter around the free energy derivative $-\beta\partial G/\partial\psi$ which is accurately reproduced by a spline approximation to the $D^{-1}\overline{\dot\psi}$ data. Integration with respect to $\psi$ of the $\overline{\dot\psi}$ data (sorted with respect to $\psi$ and linearly interpolated) thus provides an estimate of the free energy $G(\psi)$, as shown in Figure 6. This is a fundamental result: multiple short (0.5 ps) replica runs projected onto coarse variables can be used to probe the underlying free energy surfaces.

We also note that the scatter of the derivative data gives direct indication for the presence of additional slow variables. We find the largest scatter near $\psi = 1$ rad where Bolhuis et al.[34] have shown that a slow "solvent" coordinate is relevant for barrier crossing.

### C. Free energy minima and saddles as stationary points

From Fig. 5, we expect that Newton-Raphson type, contraction mapping algorithms for finding stationary points should indeed converge to the free energy maxima and minima. We test this for the most unstable saddle near $\psi_{max} = -2.15$ rad (as determined from umbrella sampling). We save ten configurations along a 5-ps MD trajectory with a tight harmonic constraint holding $\psi$ near $\psi_0 = -\pi/2$. From each of the ten configurations, we start ten $\tau = 0.5$-ps replica simulations with different velocities. Figure 7 shows the difference $\overline{\psi(\tau; \psi_0)} - \psi_0$ as a function of the average $[\overline{\psi(\tau; \psi_0)} + \psi_0]/2$. Despite the scatter in the difference data of round 1, a linear fit already predicts zero drift at $\psi_1 \approx -1.92$ rad close to the free energy maximum $\psi_{max}$. After a second round initialized near $\psi_1$, a quadratic fit (as shown in Figure 7) yields $\psi_2 = -1.98$ rad. Inclusion of data from a third round initiated near $\psi_2$ does not change the fit and predicts an unstable $(\partial \overline{\dot{\psi}}/\partial \psi > 0)$ stationary point at $\psi = -2$ rad. The rapid convergence of the Newton-Raphson-type search shows how locations of the dominant features of free energy surfaces (i.e., minima and saddles) can be identified dynamically by searching (given good initial guesses!) for stationary points of the coarse dynamics through contraction mappings.

### D. Coarse projective integration

We illustrate coarse projective integration in the two-dimensional Ramachandran plane of dihedral angles $\varphi$ and $\psi$. In each integration step we start from identical structure but random initial velocities assigned according to a Maxwell-Boltzmann distribution. The forward differences of the dihedral angles, $\overline{\varphi(t; \varphi_0, \psi_0)} - \varphi_0$ and $\overline{\psi(t; \varphi_0, \psi_0)} - \psi_0$, are calculated from 0.5-ps MD simulations of 50 replicas. Linear extrapolation is then used to project forward in time for 0.5 ps. "Lifting" is accomplished through a 0.5-ps MD run with harmonic constraints holding both $\varphi$ and $\psi$ near their new target values.

Before we illustrate the results, we should point out that this is not a particularly "good" problem for coarse integration; indeed, for these conditions the average solution very quickly (within 2 ps) finds its way to the bottom of the well. Coarse projective integration will probably be beneficial in situations in which the "healing period" is short compared to the slow dynamics (i.e., if there exists a gap in the eigenvalues of the linearization of the drift part of the problem). If a large such gap exists (i.e., if there exists a long slow transient of the coarse behavior towards the stable minimum), then projective integration has the potential to accelerate the CMD convergence there. Here, both processes – healing and drift of the expectation down into the well bottom – are relatively fast, and (for the number of copies we use) quite noisy. Figure 8 shows the resulting dynamics. We confirm that within three forward integration steps, the system reaches the free energy minimum in the $\varphi$-$\psi$ plane. Coarse projective integration can thus be exploited to potentially accelerate CMD towards stable stationary points forward in time. The reason for including the forward in time coarse projective integration in this particular case, however, is more for completeness, and to motivate the next section.

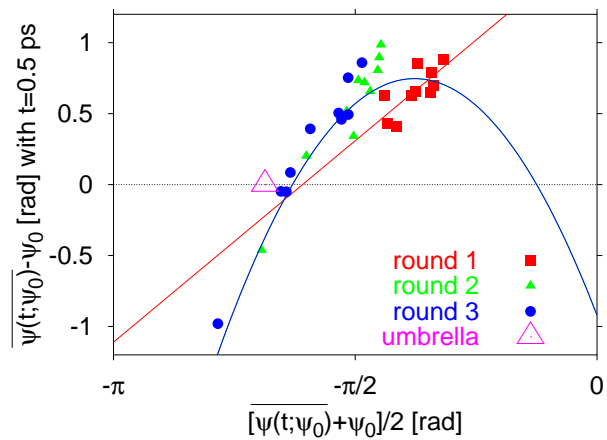

FIG. 7: Newton-Raphson search for the highest barrier near $\psi \approx -2.15$ rad ($\psi \approx -123$ deg). Starting from an initial structure with $\psi \approx -\pi/2$ rad, linear (round 1, red squares and line) and quadratic fits to $\overline{\psi(t; \psi_0)} - \psi_0$ data (round 2, filled green triangles; round 3, blue circles and parabola) are used to locate the unstable stationary point with $\overline{\psi(t; \psi_0)} - \psi_0$. The open magenta triangle corresponds to the maximum in the free energy surface, as determined by umbrella sampling.

### E. Reverse coarse integration

To escape from a free energy minimum we can also try to use reverse integration. Whereas forward integration of the expected coarse variables [here: $\overline{\psi(t; \psi_0)}$] converges towards *stable* stationary points (i.e., free energy minima), reverse integration goes "up the mountains" towards *source*-type, unstable stationary points. For a one-dimensional coarse problem the "saddle" is indeed such a "source." We start from a configuration with $\psi \approx 0$ near the bottom of the $\alpha$-helical well and perform 13 reverse integration steps of length $\Delta t = 0.5$ ps. At each step, we initialize 50 replicas with identical structure and random Maxwell-Boltzmann velocities, and run regular MD *forward* in time for $\Delta t = 0.5$ ps. We then use Eq. (7) to estimate $\overline{\psi(-\Delta t; \psi_0)}$ and "lift" the $t = 0$ structure by running a 0.5-ps MD with a harmonic constraint on $\psi$ with $\overline{\psi(-\Delta t; \psi_0)}$ as the target value. The final configuration is then used as a starting structure for the forward replica runs in the next reverse integration step.

Figure 9 shows the $\psi$ values (horizontal axis) as a function of time (vertical axis; right-hand scale) for the 13 reverse integration steps. Also shown is the free-energy surface (left-hand scale). We find that the system rapidly escapes from the free-energy minimum and reaches the barrier near $\psi = 1$ rad after three reverse integration steps ($-1.5$ ps). The subsequent fluctuations of $\overline{\psi(-t; \psi_0)}$ about the free-energy maximum are caused by statistical uncertainties in evaluating $\overline{\psi(t; \psi_0)} - \psi_0$, and by the possible role of an additional slow variable, as inferred from the study of Bolhuis et al.[34]

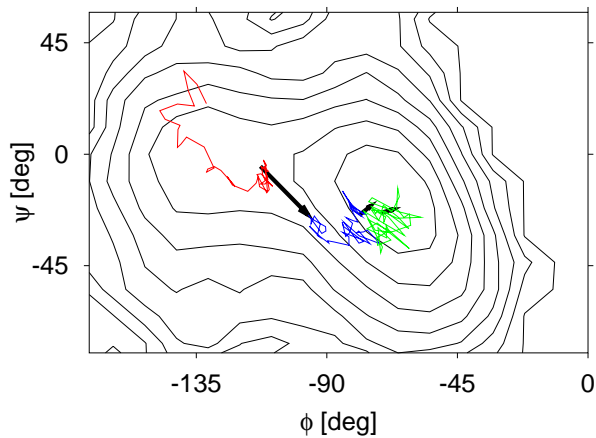

FIG. 8: Coarse projective integration in the $\varphi$-$\psi$ plane. Colored lines correspond to the average trajectories ($\overline{\varphi}$, $\overline{\psi}$). Arrows indicate the projective integration steps. Thin black lines are contour lines at $1\,\mathrm{k_B}T$ intervals.

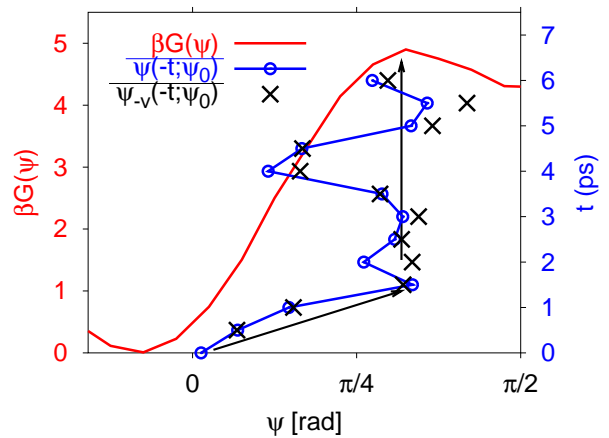

FIG. 9: Reverse integration. The blue line with open circles shows $\overline{\psi(-t; \psi_0)}$ (horizontal axis) as a function of time $t$ (vertical axis, right-hand scale). The red curve shows the free energy surface (in $k_B T$ units). The tilted and vertical black arrows indicate the initial approach to the unstable stationary point, and the subsequent fluctuations about it, respectively. Shown as black crosses are values for $\overline{\psi(-t; \psi_0)}$ obtained from reverse integration using the same starting configuration but initial velocities of opposite sign.

The time reversibility of classical mechanics is the basis for one of the main objections to Boltzmann's kinetic theory, the so-called "irreversibility paradox." Here, one might naively expect that if we reverse each of the replica trajectories by changing the sign of the initial velocities, our *reverse* integration scheme would turn into a *forward* integration. This is not the case. For each of the initial structures along the reverse-integration path found before, we run each of the 50 replicas forward in time with initial velocities of opposite sign. The resulting reverse $\psi$ values are also shown in Figure 9, and are found to agree well with those obtained for the original initial velocities, illustrating the "irreversible" time evolution of the averaged coarse variables. To qualify this result, we point out that the integrator used in the MD simulations is not fully time reversible because of thermostatting and the bond constraint algorithm.

While reverse integration will take us "up" the free-energy surface, it will eventually search for "mountain tops" rather than saddles. Even so, it can still be rationally used as a tool to help explore the free energy surface. For a coarsely two-dimensional surface, techniques for approximating the two-dimensional stable manifolds of fixed points in dynamical systems can be used to efficiently draw the surface by reverse integration. In effect, a circle of points surrounding a well-bottom gives a one-parameter family of initial conditions for reverse coarse projective integration that can be used to "triangulate" the surface. Computational approaches to these problems for explicit ordinary differential equations are well developed[49,50] and we expect that they can find good use in the case of "coarse stable manifolds." It is also worth mentioning that, if close to a saddle a large separation of time scales exists between a slow unstable mode and many fast stable ones, projective backward integration may indeed approach the saddle.

### F. Kinetics of interconversion

To estimate the rate of escape from the $\alpha$ helical minimum to the extended minimum, we apply the Chapman-Kolmogorov relation, Eq. (9). The propagators are constructed from 56 of the runs of 50 replicas used before to estimate the free energy derivative. We use a simple linear interpolation of the propagators, Eq. (12), that ignores a small translational correction for the narrowly spaced starting points $\psi_0$. Figure 6 compares the free energy surface predicted from Chapman-Kolmogorov iteration, $\beta G(\psi) = -\ln p(\psi, t \to \infty | \psi_0, 0)$, to the free energy from the equilibrium run. The agreement is excellent, with the possible exception of the poorly sampled barrier near $\psi = -2.15$ rad. This shows that Chapman-Kolmogorov iterations are indeed applicable, at least to estimate the equilibrium distribution.

To test the applicability to kinetics calculations, we determine the rate of escape from the lowest-free energy well at $\psi_0 = -0.3$ rad into the extended well with $2\,\mathrm{rad} < \psi < 2\pi - 2.5\,\mathrm{rad}$ which defines the absorbing region $[h(\psi) = 0]$ in the iteration, Eq. (9). By integration over $\psi$, we find that the survival time distribution rapidly becomes exponential with a time constant of 920 ps (see Figure 10). As a reference, we also determine the corresponding survival time distribution from the two equilibrium runs. From the 7-ns and 24-ns runs, we estimate mean-first-passage times of about 400 ps and 800 ps, respectively. The rate constant for escaping from the alpha-helical well to the extended well calculated by Chapman-Kolmogorov iterations is thus within about a factor of two of the value from simulation.

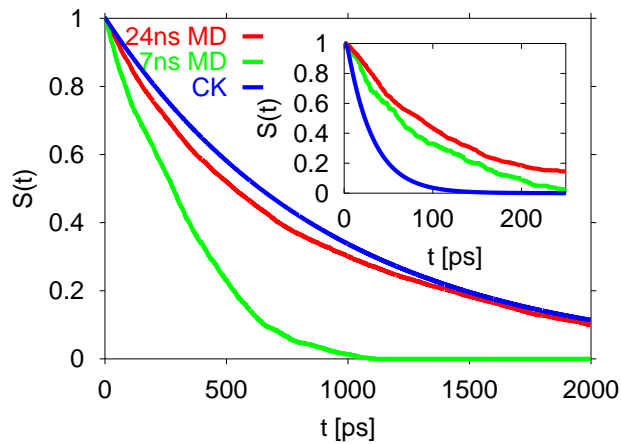

FIG. 10: Survival time distribution to reach the extended state ($2$ rad $< \psi < 2\pi - 2.5$ rad) from the $\alpha$-helical well, starting from $\psi_0 = -0.3$. Results are shown for the two equilibrium runs with 265 (red line) and 607 water molecules (green line), and for Chapman-Kolmogorov iterations (blue line). The inset shows the survival time distributions to reach the $\alpha$-helical minimum from an extended configuration ($\psi \approx \pi$).

For the backward rate coefficient to go from the extended minimum ($\psi \approx \pi$) to the $\alpha$-helical minimum ($-0.6$ rad $< \psi < 0$), the Chapman-Kolmogorov iteration gives a rate coefficient of about $1/(33$ ps), consistent with the free energy difference of about $3.3$ $k_B T$ (Figure 3) and the forward rate coefficient of $1/(920$ ps). Both MD simulations give backward rate coefficients of about $1/(100$ ps), consistent with the free energy difference of about $2$ $k_B T$ and a forward rate coefficient of $1/(800$ ps) for the 24-ns run, but too slow for the forward rate of $1/(400$ ps) estimated for the 7-ns run. We have confirmed these rate coefficients and the underlying two-state model by determining the decay time of the number correlation function, $\langle \theta[\psi(t)]\theta[\psi(0)]\rangle$ [with $\theta(\psi) = 1$ inside one well and 0 outside] and relative populations in the two wells. In the 24-ns simulation, the number correlation decays exponentially over a broad range. The decay time is insensitive to the particular choice of dividing lines in the barrier regions, and a two state model is consistent with the above rates and the observed equilibrium coefficient.

For comparison, we also calculate the rates of interconversion by numerical solution[60] of the Smoluchowski diffusion equation along $\psi$. With a diffusion coefficient $D = 0.15$ rad$^2$ ps$^{-1}$ and a free energy surface $G(\psi)$ from the equilibrium simulations, we obtain forward and backward rate coefficients of $1/(600$ ps) and $1/(120$ ps), respectively. These values are in good agreement with the rate coefficients from the MD simulations and from the Chapman-Kolmogorov iterations.

Our analysis demonstrates how CMD can use subpicosecond dynamics to extrapolate by three orders of magnitude to nanosecond dynamics. One concern may be that in a system with many wells, it will be impossible *a priori* to sample all relevant space in the coarse variables. With re-verse integration and Newton-Raphson search, however, we can create propagators for ensembles of configurations that connect between neighboring wells. Starting from different initial structures in a given well, the search will lead to multiple exit routes from that well. The Chapman-Kolmogorov iteration for this set of local propagators can then be used to project forward in time and carry the system over barriers.

## V. CONCLUSIONS

The CMD approach uses methods closely related to those of other approaches aiming at the long time dynamics. In constrained dynamics,[43] a single "coarse" coordinate is held fixed in time and the corresponding mean force is evaluated. In CMD, we evaluate instead the drift along a coarse variable resulting from the interactions with the rest of the system. In Voter's parallel replica method,[7] multiple replicas start from the same configuration with different velocities. After a first "transition" occurs in one of the replica simulations, as detected by observing "coarse" variables, all replicas are moved forward to the new configuration and reinitialized. Here, we use multiple replicas to determine the short-time dynamics in the coarse variables. This allows us to *construct* paths out of free-energy wells by recursive application of the Chapman-Kolmogorov identity, by Newton-Raphson search for saddle points, or, sometimes, by reverse integration. In the former approach, information about the rare events is contained in the tails of the propagators constructed from MD, similar to the advancing replica in Voter's parallel replica method.[7]

In the transition path sampling approach developed by Chandler, Dellago and co-workers,[6,61] extending earlier work of Pratt,[62] dynamic paths connecting reactant and product regions are efficiently sampled with weights given by an appropriate action functional. This offers a direct and elegant route to the dynamics of rare events. In CMD, we can use local short-time propagators to build transition paths in the space of the coarse variables. With the Chapman-Kolmogorov approach, we can estimate the kinetics of barrier crossing. In addition, we can sometimes exploit reverse integration to *construct* multiple coarse transition paths.

In the case of a single dominant "coarse path," deterministic optimization methods (preferably derivative-free algorithms) may be wrapped around the coarse timestepper to locate the path.

In the context of peptide folding and polymer dynamics, Kostov and Freed[63] have been extending the projection approach of Bixon and Zwanzig[64] to build Langevin dynamics models[65] for peptide dynamics. In a mode-coupling approach, the initially chosen bond-vector basis set was expanded by adding products of modes, leading to an increasingly accurate representation of the dynamics. Here, we avoid an explicit construction of the generalized Langevin equations for the coarse variables. Instead, we use "coarse timestepping" to close the evolution equations for the coarse variables. Mode-coupling approaches, as discussed by Kostov and Freed,[63] can be used to expand the set of coarse variables in CMD if necessary. To explore conformation space, Huber and van Gunsteren[66] have developed a method that couples the dynamics of multiple replicas to their average structure. These authors point out that "the average structure of a swarm of molecules converges faster to the structure with lowest energy than individual molecules do." In adapted form, this is an essential element of the CMD approach to mapping the coarse free energy surface. In the Newton-Raphson search and projective forward integration of the coarse variables, we use the time evolution of the averaged coarse variables to converge rapidly to *free energy* minima. In an earlier paper, Huber et al.[67] have explored the idea of adding memory to molecular dynamics to enhance conformational sampling. In a recent paper, Laio and Parrinello expanded on this approach and ingeniously combined coarse timestepping with "building in" a memory that allows the simulation to explore the free energy surface by, in effect, filling up the wells.[9] In their approach, repulsive markers are left behind along a trajectory projected into the coarse space. Eventually, these markers drive the system over a barrier leading out of the well (and, in the process, effectively mapping the well out). Our "stable manifold through reverse integration" maps out the wells by essentially reversing the deterministic part of the coarse dynamics.

We have shown here that CMD can be used for rapid searches of the conformation space of flexible molecules in aqueous solution. We demonstrate how the results of multiple short (0.5 ps) replica runs can be combined to determine the free energy surface using two different methods. We have illustrated how projective forward and reverse integration can be used to move toward free energy minima and saddles. We have also shown how dynamics at the sub-picosecond timescale can be used to predict the slow (0.5 to 1 ns) kinetics of barrier crossing.

These encouraging results show that CMD is a robust method. From the work of Bolhuis et al.[34] we know that additional "solvent" coordinates are kinetically relevant. The subspace monitored here does thus not cover the slow dynamics completely. Similar behavior is expected in many practical applications. While CMD is very flexible with respect to the inclusion of additional variables, the construction of slow (or "hydrodynamic") variables is often difficult. Even for the Ala dipeptide, Bolhuis et al.[34] could only identify the second relevant variable in vacuum, but not in water. Here, we show that CMD can give promising quantitative results even if only part of the slow dynamics is covered. An explanation for this result is the averaging invoked here. By starting from configurations with similar $\psi$ values, but different drift directions (or commitment probabilities), we average over additional slow variables. Moreover, the presence of such variables is apparent from the data as a large scatter in the drift velocities and directions. Such scatter highlights the need to either add new variables or perform additional simulations.[18]

In summary, the CMD approach provides an integrated framework, individual components of which are closely related to various approaches aimed at overcoming the time-scale problem in MD. Here, we have demonstrated that CMD is a useful approach to extract thermodynamic and kinetic properties of molecular systems, and to extrapolate their long-time dynamics.

One of the ambitions of coarse computation is not only to map "coarse phase space" effectively, but also to search parameter space efficiently. Indeed, coarse bifurcation algorithms can be implemented based on the coarse timestepper approach that *converge on* parameter values at which qualitative transitions occur for the coarse dynamics (see, for example,[18,20,22]). In the present context, this would correspond to finding the regime of parameters like temperature or solution ionic strength at which coarse free energy wells form. Such coarse timestepper based numerical bifurcation and continuation techniques have been demonstrated for kinetic Monte Carlo, Lattice-Boltzmann and Brownian Dynamics cases; we are currently transferring this computational technology to the CMD context. Finally, the CMD approach is not limited to dynamics on a classical energy surface and can be implemented equivalently if the dynamics occurs on a quantum surface as, for instance, in Car-Parrinello MD.[68] In this context, it is conceivable that the concept of "telescopic projective integration" for systems with multiple gaps in their eigenvalue spectrum[40] may find a fruitful application.

### Acknowledgments

G.H. wants to thank A. Szabo and R. Zwanzig for many insights and comments. Helpful discussions with several individuals, notably D. Barkley, J. Carey, D. Maroudas, A. Z. Panagiotopoulos, S. Shvartsman, A. Stuart are gratefully acknowledged. I.G.K. would like to acknowledge the long-standing collaboration with C. W. Gear in the development of equation-free methods. This work was partially supported by AFOSR (Dynamics and Control) and an NSF-ITR grant (I.G.K.).